\documentclass[conference]{IEEEtran}
\usepackage{graphicx} 
\usepackage{cite}
\usepackage{colortbl}

\newcommand{\linebreakand}{%
  \end{@IEEEauthorhalign}
  \hfill\mbox{}\par
  \mbox{}\hfill\begin{@IEEEauthorhalign}
}

\title{DTRAN: A Special Use Case of RAN Optimization using Digital Twin}

\author{\IEEEauthorblockN{Caglar Tunc\IEEEauthorrefmark{1},
		Kubra Duran\IEEEauthorrefmark{2}\IEEEauthorrefmark{4}, Buse Bilgin\IEEEauthorrefmark{3}, Gökhan Kalem\IEEEauthorrefmark{3}, Berk Canberk\IEEEauthorrefmark{2}\IEEEauthorrefmark{5}}
	\IEEEauthorblockA{\IEEEauthorrefmark{1}Next-Gen R\&D/6GEN Lab, Turkcell, İstanbul, Türkiye}
 
 \IEEEauthorblockA{\IEEEauthorrefmark{2}School of Computing, Engineering and The Built Environment, Edinburgh Napier University, UK}

\IEEEauthorblockA{\IEEEauthorrefmark{3}Next-Gen R\&D, Turkcell Technology, İstanbul, Türkiye}

\IEEEauthorblockA{\IEEEauthorrefmark{4}BTS Group, İstanbul, Türkiye}
\IEEEauthorblockA{\IEEEauthorrefmark{5}Department of Artificial Intelligence and Data Engineering, Istanbul Technical University, Türkiye}

		Email: \IEEEauthorrefmark{1}caglar.tunc@turkcell.com.tr, \IEEEauthorrefmark{2}\{kubra.duran, b.canberk\}@napier.ac.uk,
		\IEEEauthorrefmark{3}\{buse.bilgin, gokhan.kalem\}@turkcell.com.tr,\\ \IEEEauthorrefmark{4}kubra.duran@btsgrp.com, \IEEEauthorrefmark{5}canberk@itu.edu.tr}

\date{March 2024}

\makeatletter

\let\old@ps@IEEEtitlepagestyle\ps@IEEEtitlepagestyle

\def\confheader#1{%
    \def\ps@IEEEtitlepagestyle{%
        \old@ps@IEEEtitlepagestyle%
        \def\@oddhead{\strut\hfill#1\hfill\strut}%
        \def\@evenhead{\strut\hfill#1\hfill\strut}%
    }%
    \ps@headings%
}
\makeatother

\confheader{%
    \parbox{20cm}{Accepted by 2024 European Conference on Networks and Communications EuCNC \& 6G Summit, \textcopyright 2024 IEEE}
}

\begin{document}

\maketitle


\begin{abstract}
The emergence of beyond 5G (B5G) and 6G networks underscores the critical role of advanced computer-aided tools, such as network digital twins (DTs), in fostering autonomous networks and ubiquitous intelligence. Existing solutions in the DT domain primarily aim to model and automate specific tasks within the network lifecycle, which lack flexibility and adaptability for fully autonomous design and management. Unlike the existing DT approaches, we propose RAN optimization using the Digital Twin (DTRAN) framework that follows a holistic approach from core to edge networks. The proposed DTRAN framework enables real-time data management and communication with the physical network, which provides a more accurate and detailed digital replica than the existing approaches. We outline the main building blocks of the DTRAN and describe the details of our specific use case, which is RAN configuration optimization, to demonstrate the applicability of the proposed framework for a real-world scenario.
\end{abstract}

\section{Introduction}

As the rollout of 5G networks and research in 6G ramp up, computer-aided solutions for network optimization and automation become increasingly important for the telecommunication ecosystem. To capture the complex nature of these networks, mobile operators and researchers adopt advanced software solutions, including Digital Twins (DTs) of the wireless network, which raises the computational power and efficiency of digital platforms beyond the conventional simulation tools \cite{tunc_2024}. The main goal for the DT platforms is to create a digital replica of the physical network, which is used for continuous analysis, testing and optimization. Algorithms and parameters can be validated and optimized first on this digital replica before interacting with the physical network, preventing the potential adverse impacts on the network performance. Then, the outputs and recommendations made by the DT environment can be fed back to the physical counterpart, paving the way for more robust operation, optimization and automation.


Current solutions for DT mainly focus on modelling specific components and systems of the network, such as core network \cite{10143356, rodrigo_2023, tao_2024}, wireless networks \cite{t6conf, wifi} or radio access network (RAN) \cite{nguyen_2021}. These solutions mainly aim to develop a digital model to optimize the system for a specific task and service \cite{khan_2022}. Hence, the flexibility and adaptability to new scenarios and use cases for fully autonomous network control is an open challenge. Furthermore, depending on existing and pre-trained models for network management \cite{wiley} hinders the current solutions' capabilities to handle the unpredictable nature of 6G networks \cite{kuruvati_2022, dtaas, apostolakis_2023, spatio}.

To fill this gap, we introduce the special use case of RAN optimization using DT (DTRAN) framework as a comprehensive solution that enhances adaptability, accuracy, and autonomy in network design and management. Unlike existing approaches, this framework provides a better understanding of the network and detailed “what-if" analyses to capture the impact of various configurations and parameters. Our main contributions and novelty over the existing efforts in the literature are given below:
\begin{itemize}
    \item We propose a more accurate and comprehensive DTRAN framework that provides the network operators with in-depth insights into network behaviour, including the effect of different network parts.
    \item The envisioned DTRAN framework enables real-time data management and communication with the physical counterpart, which makes it a good fit for the long-term vision for fully autonomous networks.
    \item Our proposed framework uses innovative real-time analytics and decision-making algorithms to proactively empower operators to identify and address emerging challenges.
\end{itemize}

\begin{figure*}[t]
    \centering    \includegraphics[width=0.8\textwidth]{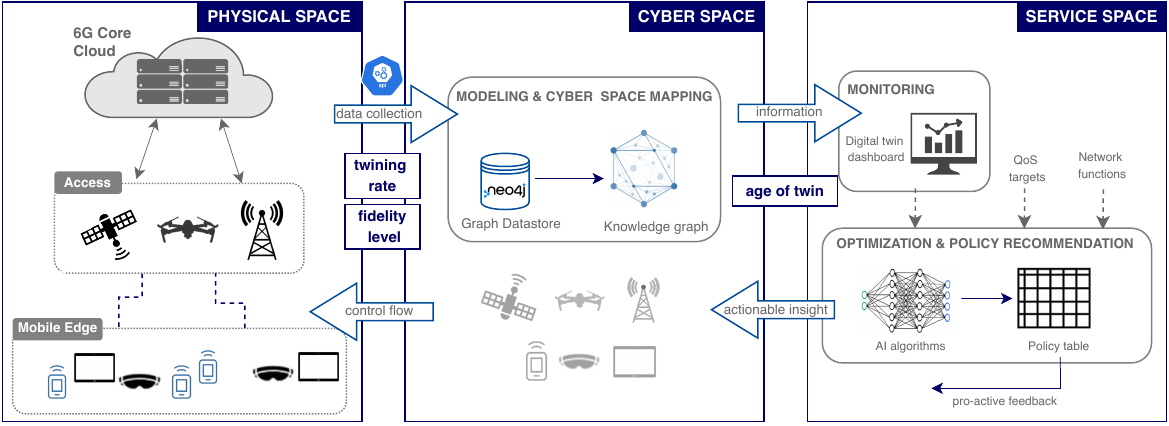}
    \caption{Proposed DTRAN framework.}
    \label{fig:dt_arch}
\end{figure*}

\section{Preliminaries}
We use several metrics to quantify the requirements and the performance of DTRAN. These are listed as follows:
\begin{itemize}
    \item \textbf{Twinning rate} is the rate at which the DTRAN is recalculated and/or calibrated to maintain accuracy with respect to the physical network.
    \item \textbf{Age of twin} is the metric that quantifies the freshness of the DTRAN model. By definition, twinning rate and the age of twin are inversely proportional, since a higher twinning rate reduces the age of twin \cite{aot}. 
    \item \textbf{Fidelity level} measures the accuracy and explainability of the DTRAN, which depends on the embedded models and the amount and frequency of data exchange. It is determined by the level of details and abstractions embedded in the model. A higher fidelity level is generally desired since it means a more accurate representation of the physical twin.  However, high fidelity level necessitates a more complex DT model and data processing, limited by the storage and computational capabilities of the physical resources.
\end{itemize}

\section{dtran Framework}
As seen in Fig. \ref{fig:dt_arch}, DTRAN consists of three spaces: \textit{(i) Physical Space} leverages real-time data streams from diverse sources, including satellites, mobile network infrastructure and ground sensors, to continuously update a dynamic and high-fidelity network infrastructure, \textit{(ii) Cyber Space} utilizes graph-based data modelling and knowledge graphs to handle multi-sourced data relations, \textit{(iii) Service Space} leverages enhanced AI models to provide optimization and policy recommendations for targeted Quality of Service (QoS) levels and network functions. 

\section{Sample Use Case - Fully-Automated RAN Configuration Optimization}
RAN configurations involve the coordination and cooperation of hundreds of parameters, such as tilt, power, handover and carrier aggregation parameters, that determine how specific network functions and RAN operates. Operating such large number of parameters and systems harmoniously can be challenging for network operators. As seen in Fig. \ref{fig:auto_config}, DTRAN model can leverage fully automated RAN configuration optimization via:


\begin{figure}[h]
    \centering    \includegraphics[width=0.43\textwidth]{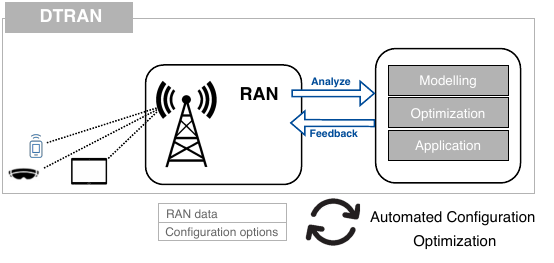}
    \caption{Automated RAN configuration optimization.}
    \label{fig:auto_config}
\end{figure}

\begin{itemize}
    \item \textbf{Data collection and modelling :} Streamlined RESTful APIs and data extraction tools facilitate RAN data retrieval to establish a DT warehouse. Then, graph-based data modelling is performed via Neo4j \cite{noauthororeditorneo4j} for RAN data to preserve the relationships and increase the model accuracy via fine-tuned twining rate. 
    \item \textbf{Monitoring and optimization:} Real-time twin models, including user equipment (UEs) and base stations (gNodeBs), are monitored on the DT dashboard to trace the dynamic changes within the RAN environment with high fidelity. DTRAN implements Deep Reinforcement Learning (DRL) algorithms to optimize configuration by setting the desired objective/utility function and constraints.
    \item \textbf{Application and feedback:} The optimized configurations are tested by running iterative scenarios to minimize communication failures and ensure stable age of twin values. Then, they are applied back to the RAN topology via a feedback channel.
\end{itemize}

\section*{Acknowledgment}
This work was supported by The Scientific and Technological Research Council of Turkey (TUBITAK) 1515 Frontier R\&D Laboratories Support Program for BTS Advanced AI Hub: BTS Autonomous Networks and Data Innovation Lab. Project 5239903.


\bibliographystyle{IEEEtran}
\bibliography{IEEEabrv, whole}

\end{document}